\documentclass[latexsym]{article}

\begin{document}

\textbf{Note: tex source modified by arXiv admin to permit rendering without the omitted style
file reta.sty. Original source is available in the orig subdirectory.}

\section*{Categorical Physics}

Louis Crane \\
        Department of Mathematics \\
        Kansas State University \\
        Manhattan, KS   66506

\subsection*{I. Introduction}

The purpose of this letter is to outline a rather novel suggestion as to
what form the quantum theory of gravity should take.  The suggestion has
a certain mathematical elegance, but that was not the motivation which
brought it into being.  Rather, it was motivated by the juxtaposition of
two ideas:

(A) the work of Rovelli and Smolin on the loop representation
(1) which showed that a state for the quantum theory of gravity could be
described by an invariant of links in a $3$-manifold, given by
integrating a measure on the space of connections on the manifold
against the traced holonomy around the link.

(B) the work of E.\ Witten (2) which recovered a generalized Jones
polynomial as precisely such an integral; using the Chern-Simons
invariant to create a measure.

The ``measures'', in A and B above are, of course, formal.  It was in
order to make mathematical sense of Witten's Chern-Simons invariant that
I showed in (3) that any modular tensor category gave rise to a  $3-D$
Topological Quantum Field Theory, i.e.\ allows us to reproduce all the
results of Witten's path integral by rigorous methods.

The juxtaposition of ideas (A) and (B) tells us that the Chern Simons
measure gives us a state for quantum gravity on a closed  $3$-manifold,
which can be reproduced entirely from an algebraic structure called a
modular tensor category, first described in (4).  The construction gives
us more than a state (i.e.\ link invariant) on a closed  $3$-manifold;
it also gives us vector valued invariants of links in  $3$-manifolds
with boundary, and a nice factorization when we join along boundary
components.  This is what we mean when we say that we can construct a
topological quantum field theory.

What I want to propose is that the universe as a whole is in a
Chern-Simons state (presumably for a larger group than  $SU(2)$, in
order to include matter - more on this latter).

The connection of the Chern Simons state to a TQFT is what allows us to
experience a world of changing phenomena in a universe which is not
changing as a whole.  Manifolds with boundary correspond to observers,
and observations take place in the Hilbert spaces which our TQFT assigns
to boundaries.  Thus, in order to quantize gravity (or gravity plus
matter) we have to abandon the idea of measurement at a distance, and
create a relative form of the probability interpretation of quantum
mechanics; with a Hilbert space for each observer, and consistency
relations which are best expressed in categorical language.  We can say
that a state for quantum gravity is given by a functor from the category
of observers to the category of vector spaces; this coincides with the
notion of a TQFT.  Time evolution must be given by a natural
transformation between functors.

This is obviously an ambitious program.  As yet, I am not able to
compute the result of any experiment in this framework, but I can see a
path which could lead to that goal.  In the rest of this letter, I
summarize what I can understand of various aspects of this program, and
suggest various directions for further research.

\vskip .2in

\subsection*{II. The Category of Observers}

In this section, I make a series of fundamental definitions.  On the one
hand, they are motivated by the mathematical juxtaposition mentioned
above; on the other, they have, I believe, a certain physical
plausibility.

Definition: {\bf An \underbar{Observer} is an oriented  $3$
manifold with boundary containing an embedded labeled framed graph which
intersects the boundary in isolated labeled points.}

(The labeling sets for the edges and vertices of the graph are finite,
and need to be chosen for all of what ensues.)

Definition: {\bf A \underbar{skin of observation} is a closed oriented
surface with labeled punctures.}

Definition: {\bf If A and B are skins of observation an
\underbar{inspection, $\alpha$, of B by A} is an observer whose boundary
is identified with  $\overline{A} \cup B$  (i.e.\ reverse the
orientation on A) such that the labelings of the components of the graph
which reach the boundary of  $\alpha$  match the labelings in
$\overline{A} \cup B$.}

This definition requires that the set of labelings possess an involution
corresponding to reversal of orientation on a surface.

To every inspection  $\alpha$  of B by A there corresponds a dual
inspection of  $\overline{A}$  by  $\overline{B}$, given by reversing
orientation and dualizing on  $\alpha$.

Definition: {\bf The \underbar{category of observation} is the category
whose objects are skins of observation and whose morphisms are
inspections.}

Definition: {\bf If  $M^3$  is a closed oriented  $3$-manifold the
\underbar{category of observation in  $M^3$} is the relative (i.e.\
embedded) version of the above.}

Of course, we can speak of observers, etc. in  $M^3$.

Definition: {\bf A state for quantum gravity (in  $M^3$)  is a functor
from the category of observation (in  $M^3$)  to the category of vector
spaces.}

Let us reiterate that all the above definitions should be extended by
specifying the labeling set with involution.

Observation: {\bf Any modular tensor category gives rise to a state for
quantum gravity.}

The mathematical motivation for the above definitions should be clear;
it is just what we know how to produce.

In our language, a State for quantum gravity is the same thing as a
$3-D$  TQFT.  We should note some facts about the development of the
Ashtekar variables for general relativity, which make this physically
plausible.  Although a state is given by any invariant of {\bf links},
since the Hamiltonian constraint only gives a condition at the vertices
of a {\bf graph}, the states which are supported only on links seem to
have unsatisfactory properties.  In searching for solutions of the
Hamiltonian constraint which were nontrivial on {\bf graphs} Gambini et
al (5) found themselves reproducing the perturbative terms of the CSW
invariant.  Thus, it could well be that CSW gives the only physically
interesting states.  Should the connection with TQFT then be taken as
pure coincidence?

Let us now consider the physical ideas which underlie these definitions.
The first departure is attaching vector (``Hilbert'') spaces to skins of
observation, rather than observers.  This means that if we divide the
universe into ``system'' and ``observer'' the observer no longer
measures the state of the system, but only of that part of it which
impinges instantaneously on the observer.  This is commonly neglected in
ordinary quantum mechanics, but it seems reasonable that in quantizing
general relativity we would need to consider it.  I refer to this as
abandoning ``measurements at a distance''.

We are left with a picture in which states for a whole universe are
fixed things constructed out of special algebraic structures, while the
relational states which we measure live in spaces constructed by a
different recipe from the same structures.

Here we see a second physical departure.  The {\bf state of the universe
as a whole} becomes part of the framework of laws by which the parts
interact.

It has been noted for some time that when considering the entire
universe, the relationship between laws and initial conditions could be
different from what physicists commonly expect.  Hawking has suggested
determining initial conditions from laws.  I am proposing, rather that
the ``initial'' state of the universe (which does not change in time)
acts as a law.

It is not hard to create a quantum mechanical system in which the
initial state of a combined system acts as a law of motion for the two
parts [10], so perhaps this notion is not so implausible.

Before discussing how this theory might meet some of the titanic
challenges which await it, let me allow myself a few philosophic
observations.

Category theory was invented largely in order to rewrite mathematics in
the most coordinate invariant way possible.  The notion that it was
desirable to write things invariantly is something which 20th century
mathematics inherited from relativity.  (The first abstract definition
of a vector space is in Weyl's exposition of relativity {\bf Zeit}, {\bf
Raum}, {\bf Materie}.)

Thus, if the fully adequate formulation of relativity uses categories,
we would be coming full circle.

Beside the functorial nature of the {\bf result}, i.e.\ of a state of
quantum gravity; there is also a categorical structure, namely a modular
tensor category, at the heart of the construction [3] of a  $3-D$  TQFT,
which we are reinterpreting here.

I want to argue that the entire CSW invariant is more interesting than
its perturbative terms.  This is because modular tensor categories are
only slight modifications of Tannakian categories.  [6] Tannakian
categories are simply the categories of representations of groups.
Thus, tensor product categories are expressions of symmetries.  Modular
tensor categories are ``purely quantum'' versions of the same thing.
Thus, we can think of them as ``quantum symmetries''.  The fact that we
can construct our invariant states from MTC's can be expressed (somewhat
lavishly) by saying that in the quantum domain invariant states are
determined by symmetry.

Enough philosophy.  How can this theoretical model be compared with
experiment?  I believe that there are three main problems which must be
confronted: (A) making some physical interpretation of the states which
we attach to skins of observation (B) the problem of introducing time
and (C) how to include matter.  In the next 3 paragraphs, I shall say
what I know about these problems.

\vskip .2in

\subsection*{III. States as Quantum Geometries.  Deformed Spin Networks}

The idea I discuss here was already proposed in [7].  I shall outline it
here for completeness, and to set it in a theoretical context.

The invariants which CSW theory assigns to embedded knotted labeled
graphs are deformed generalizations of the number called the
``evaluation'', which Penrose [8] assigned to a labeled graph called a
spin network.

Regge and Ponzano [9], were able to interpret the evaluation as a sort
of discrete path integral for  $3-d$  euclidean quantum gravity.  The
formula which Regge and Ponzano found for the evaluation of a graph
$$
\# = \sum_{\textrm{labelings}} \ \prod_{\textrm{int. edges}}\  qdimJ\
 \prod_{\textrm{Tetrachedra}} \ \{ 6J\},
\hfill 1
$$
follows from some elementary properties of representation theory.  In
(1), we have placed our trivalent labeled graph on the boundary of a
$3$-manifold with boundary, then cut the interior up into tetrahedra,
labeling the edges of all the internal lines with arbitrary spins.  (The
Clebsch-Gordon Relations imply that only finitely many terms in (1) are
non-zero.)

The evaluation of a tetrahedron for a spin network is a  $6J$  symbol
for the group  $SU(2)$.

The derivation of formula (1) uses only basic properties of the category
of representations of  $SU(2)$.  Thus, we can repeat the derivation of
(1) for CSW invariants of trivalent graphs, provided that we are careful
not to embed the graph in such a way that we have to unbraid it in order
to put it on a boundary.  This is not an issue for spin networks, since
over and under crossings are equivalent (i.e.\ the tensor category is
symmetric).

There is a natural generalization of spin networks to larger groups,
called {\bf fabrics} by Moussouris [11].  A formula like (1) can be
similarly deduced, except that the ``labelings'' include assigning
intertwining operators to internal faces.  (These are unique for
$SU(2)$.)

Thus, a formula like (1) but with more general labelings, including
labels on faces, can be derived for CSW invariants of graphs.

Formula (1) has the same form as the invariant of  $3$-manifolds due to
Viro and Turaev, but the geometric setting is different, and we are
obtaining the CSW invariant, instead of its absolute square.

Regge and Ponzano reinterpreted (1) as a discrete path integral for
$3-d$  quantum gravity.

The essential step for them was to reinterpret the spin label on each
edge as a length.  Assigning lengths to the edges of a triangulation is
an approximation to selecting a metric; summing over all labelings then
approximates a path integral over metrics.

Thus a labeling of the edges of a graph gives a sort of discretized
probability density of metrics or ``quantum geometry'', to the space it
surrounds.  For ordinary spin networks, these sums peak around flat
metrics, so the classical ``equations of motion'' match those for  $3-d$
euclidean general relativity.

Applying the same approach to states on the vector space attached to a
boundary surface yields a ``quantum geometry'' on {\bf part} of the
interior of a  $3$-manifold with boundary.  A small improvement of the
form of (1) would suffice to give a quantum geometry on the entire
interior.  Studying the recursion relations for quantum  $6J$  symbols
leads me to believe the improvement exists, but I have not constructed
it, and do no pursue it here.  At any rate, being able to interpret a
state on the boundary as probabilistic information about the geometry on
the interior brings us much closer to relating our formalism to
experiment.  If we could relate states on a boundary for CSW theory with
a larger group to a more complex geometry in the interior, mixing
intrinsic with extrinsic geometry (i.e.\ coupling gravity to Yang-Mills)
it would be even a larger step.

This brings us to the question of introducing matter into our picture.

\vskip .2in

\subsection*{IV. Matter and Larger Groups}

The  $2 + 1 - D$  TQFT constructed by means of a modular tensor category
(or quantum group, etc.), can equally well be defined for some larger
semisimple compact lie group as for  $SU(2)$.  [Noncompact forms are
much thornier.]  It is tempting (to anyone who finds this paper
tempting), to try to interpret them as ``states of gravity plus
matter.''  This raises a key issue.

Issue: {\bf Are TQFT's associated to Larger Groups States for some
Quantum Theory?}

If the answer to this question is yes, then it is not necessary to worry
too much about finding the Yang-Mills Lagrangian in the theory, since
the action of the renormalization group will bring us to it if our
theory has the right symmetries.

As it turns out, it is possible to approach this problem, although not
in a mathematically rigorous way.  Peldan [12] has written down a
Hamiltonian constraint for an arbitrary lie group, which reduces to the
Ashtekar form for  $SU(2)$, and produces no second class constraints.
We can then translate our question into the simpler question of whether
the Chern-Simons State  $\ell^{2\pi ik cs(A)}$  solves Peldan's
generalized
Ashtekar Hamiltonian constraint (with cosmological constant).  This
reduces to some algebraic relations on the group, currently under study.
This would not, of course, provide a rigorous argument that the TQFT
associated to some special group is a quantum state for
gravity-plus-matter, but it would be very suggestive.

\vskip .2in

\subsection*{V. The Problem of Time}

The description of the world which the picture we are discussing
provides would not, (even given the most optimistic outcome of the
unsettled questions) at this point, resemble the world we see.  This is
because the relative states which live on surfaces have no time
evolution.  Past, present and future are all one in them.

This is not a new problem in attempts to quantize gravity.  It is a
consequence of the vanishing of the Hamiltonian on physical states,
which is ultimately an expression of  $4-D$  diffeomorphism invariance.

There is a standard resolution of this question among relativists.  This
is to treat one part of the variables as a clock.  The relative states
of the rest of the variables then turn out to solve a Schr\"odinger
equation [13].

Thus, if we are to interpret the states in our model as giving an
evolution in  $4-D$  space time, we have to solve the problem of
representing clocks within the states of a TQFT.

There is another fundamental question which must be faced here: Can one
have a clock in a purely gravitational universe, or are clocks
necessarily material?  As Einstein once phrased a similar question: If
matter disappeared from the universe, would time as well?

Moncrief [14] has taken rather heroic measures, to construct a clock from
a gravitational universe.  However, it is hard to imagine doing general
relativity with only one clock, and measuring the volume of the universe
has little to do with what an observer timing some event normally does.
I therefore believe that the most likely route to an interpretation of
the picture we are developing must go through interpreting a larger
group as gravity plus matter, and constructing clocks from matter.

Although it is hard to construct a state involving time explicitly, it
is easy to see what mathematical form such a thing should take.  We
formulate such a picture, then make some speculations as to how it might
be constructed.

Definition: {\bf A \underbar{process} is an oriented $4$-manifold with
boundary with an embedded labeled branched surface which meets the
boundary transversely in a labeled knotted graph.}

Definition: {\bf An observer with no boundary is a \underbar{moment}.}

Definition: {\bf If  $A$  and  $B$  are moments and  $P$  is a process
with boundary  $\overline{A} \cup B$, then we say  $P$  is an
\underbar{intervening process} between  $A$  and  $B$.}

If  $P$  is an intervening process between  $A$  and  $B$, then
$\overline{P}$  is an intervening process between  $\overline{B}$  and
$\overline{A}$  (``time reversal'').

Now let us define the structure which seems natural to express time
evolution.  Suppose given a vector field  $\phi$  which is the gradient
of a morse function on  ${\mathcal P}$  which intersects  $\overline{A}$
and  ${\mathcal B}$  transversely, pointing ``inward'' on  $\overline{A}$
and ``outward'' on  ${\mathcal B}$.  Dragging along  $\phi$  then gives a
functor  $F_\phi$  from the relative category of observation  ${\mathcal
O}_A$  to  ${\mathcal O}_B$.

Definition: {\bf We say that a state of gravity  $S$  is
\underbar{augmented} if for every two moments  $A$, $B$, process  $P$
intervening between them, and vector field  $\phi$  as above, we are
given a natural transformation  ${\mathcal N}_{P,\phi} : S_A \to S_B \circ
F_\phi$.  We require the assignment to be natural, in the sense that the
composition of processes and vector fields (obvious definition) is
assigned the composition of natural transformations.}

Let us translate into more prosaic language.  A natural transformation
is given by maps between the images of objects, which make certain
squares commute.  These maps are between the Hilbert spaces of skins of
observation and their ``time evolutions'', i.e. draggings by  $\phi$,
and should be thought of as giving time evolution for relative states.
The commuting squares give a consistency relation, in which the state
$M$  observes in  $N$  changes to the same state in  $F_\phi (M)$  as
the state the later observer  $F_\phi (M)$  would observe from the
changed state in  $F_\phi (N)$, i.e. the following diagram commutes:
\pagebreak

\vspace*{1.5in}

        (figure 1)

Conjecture: {\bf The natural world is described by an augmented state of
quantum gravity.}

We should note that even if we could construct an augmented state, the
physical interpretation of it would still involve all the subtleties of
clocks, lapses, shifts, etc. with which general relativity is so richly
endowed.

\vskip .2in

\subsection*{VI. Some Observations on Augmented States}

The suggestion here is that quantum gravity is to be understood, {\bf
not} as a  $3 + 1 - D$  TQFT, but as a  $(2 + 1) - D$ TQFT with some
extension into four dimensions.  The extension is very ``categorical'',
a  $2 + 1 - D$  TQFT is a set of functors, we now need functors plus
natural transformations.

This is reminiscent of the conjecture, [3], since demonstrated by K.\
Walker [15], that the  $3 - D$  TQFT's constructed from modular tensor
categories related to WZW models can be interpreted as containing
information about a bounding  $4$-manifold, (thus removing the phase
ambiguity).

More recently, I.\ Frenkel and the present author [16] have been able to
associate to a quantum group a tensor  $2$-category, which is a sort of
``recategorification'' of a modular tensor category.  The relationship
between augmented states and  $2$-categories is very suggestive.  For
those not versed in category theory, let me remark that both (skins of
observation, inspections, processes with corners) and (categories,
functors, natural transformations) are  $2$-categories [17].

\vskip .2in

\subsection*{VII. Possible Approximations}

How could we compute anything in this picture?  One suggestion, due to
L.\ Smolin, is to try to model an asymptotically flat situation.

We could imagine picking a particular weave to model the state in the
flat region, and couple it to our deformed spin networks in the
interior.

In general, it is possible that experiments could be modeled after some
hybrid of the loop states [18] in one half of the universe with the TQFT
state on the other.  This would correspond to breaking gauge invariance
in the ``apparatus''.

It may be possible to formulate a simplified model of matter by using
the representations of an inhomogeneous quantum group instead of a
larger simple group.

\vskip .2in

\subsection*{VIII. Perspective and Apologia}

One way of looking at the proposal in this paper is that it is an
attempt to fuse the many worlds interpretation of quantum mechanics with
general relativity.  States become correlated only locally, i.e. only on
skins.  Somehow, the classical equation of motion must emerge in a
situation where many bodies measure one another's state very often.

The great hope of the subject would be that once we introduce clocks
Einstein's equations in  $4 - D$  emerge by a process analogous to the
emergence of flat  $3$-geometries in the work of Regge and Ponzano.

Finally, an apologia.  This is a very sweeping proposal, motivated
partly by work in quantum gravity, but also largely by mathematical
elegance, and the existence of certain elegant algebraic structures.  It
has no doubt escaped no ones attention that much of the necessary
delicate analysis is not done.  Nevertheless, my intuitive feeling is
that enough of the ingredients of nature have mathematical analogs here
to make the ideas worth pursuing.  Mathematical elegance has served
physics well enough in the past.

\subsection*{References}

1.C. Rovelli and L. Smolin, Loop Space Representations of Quantum
General Relativity, preprint IV 20 Phys. Dept. V. di Roma ``la
Sapienza'' 1988

2. E. Witten, Quantum Field Theory and The Jones Polynomial, IAS
preprint
HEP 88/33

3. L. Crane, 2-d Physics and 3-d Topology, Commun. Math. Phys.
135 615-640 (1991)

4. G. Moore and N. Seiberg, Classical and Quantum Conformal Field
Theory, Commun Math. Phys. 123 177-254 (1989)

5. B. Brugman, R Gambini and J. Pullin, Gen. Rel. Grav. in press

6. L. Crane, Quantum Symmetry, Link Invariants and Quantum Geometry,
in Proceedings XX International Conference on differential Geometric
Methods in Theoretical Physics Baruch College (1991)

7. L. Crane, Conformal Field Theory, Spin Geometry, and Quantum Gravity,
Phys.  Lett. B v259 \#3 (1991)

8. R. Penrose, Angular Momentum; an Approach to Combinatorial Space
Time, in Quantum Theory and Beyond ed T.Bastin(Cambridge)

9. G. Ponzano and T. Regge, Semiclassical Limits of Racah Coefficients,
in Spectroscopic and Group Theoretical Methods in Physics, ed. F.
Bloch (North-Holland, Amsterdam)

10. C. Rovelli, Class. Quan. Grav. 8 297 (1991)

11. J. Moussouris, Quantum Models of Space Time Based on Recoupling
Theory, Thesis, Oxford University (1983)

12. P. Peldan, Phys. Rev. D46 R2279 (1992)

13. L. Smolin, Personal communication

14. V. Moncrief, personal communication

15. K. Walker, On Wittens 3 Manifold Invariants, unpublished

16. L. Crane and Igor Frenkel, Hopf Categories and Their
Representations, to appear.

17. M. Kapranov and V. Voevodsky, Braided Monoidal 2-Categories,
2-Vector Spaces and Zamolodchikov's Tetrahedra Equation,
preprint

18.C. Rovelli and L. Smolin, Phys. Rev. Lett. 61 1155 (1988)

\end{document}